\documentclass[sigconf,screen]{acmart}

\setcopyright{none}
\renewcommand\footnotetextcopyrightpermission[1]{} 
\AtBeginDocument{%
  }

\usepackage[utf8]{inputenc}
\usepackage{url}
\usepackage{colortbl}
\usepackage{balance}
\usepackage{xspace}
\usepackage{verbatim}
\usepackage{bold-extra}
\usepackage{amsmath}
\usepackage{multirow}
\usepackage{listings}
\usepackage{boxedminipage}
\usepackage{soul}
\usepackage{enumitem}
\usepackage[normalem]{ulem}
\setlist{nolistsep,leftmargin=.5cm}
\usepackage{booktabs}
\usepackage{tabularx}
\usepackage{svg}
\usepackage{calc}
\usepackage{float}
\usepackage{multirow}
\usepackage[normalem]{ulem}
\useunder{\uline}{\ul}{}
\usepackage{amsmath}
\usepackage[most]{tcolorbox}
\definecolor{MidnightBlue}{HTML}{006895}
\definecolor{BoxesBlue}{HTML}{DEECFF}
\definecolor{BoxesYellow}{HTML}{FFF2CC}
\definecolor{StateGreen}{HTML}{91C788}
\definecolor{StateRed}{HTML}{FF8080}
\definecolor{ArrowGreen}{HTML}{61B15A}
\definecolor{ArrowViolet}{HTML}{BA94D1}
\usepackage{microtype} 
\usepackage{algorithmic}
\usepackage{graphicx}
\usepackage{textcomp}
\usepackage{xcolor}
\usepackage{ifthen}
\usepackage{wrapfig}
\usepackage{cleveref} 
\usepackage{enumitem}
\usepackage{indentfirst}
\usepackage[T1]{fontenc}

\newboolean{showcomments}
\setboolean{showcomments}{false}

\newboolean{showboxes}
\setboolean{showboxes}{true}

\usepackage{adjustbox}
\usepackage{totcount}

\ifthenelse{\boolean{showcomments}}

\setboolean{showcomments}{true}

\newcommand{\ie}{\textit{i.e.},\xspace}
\newcommand{\eg}{\textit{e.g.},\xspace}

\usepackage{tikz}

\definecolor{bug_red}{rgb}{.84,.23,.29}

\definecolor{info-needed-color}{rgb}{1,.8,.12}

\newcounter{findingcounter}
\newtotcounter{totalfindings}
\ifthenelse{\boolean{showboxes}}
{
    \newcommand{\finding}[1]{%
      \refstepcounter{findingcounter}
      \begin{tcolorbox}[boxsep=1pt,left=2pt,right=2pt,top=1pt,bottom=1pt]%
      \small
      \centering
      \textbf{Finding \arabic{findingcounter}:} #1
      \end{tcolorbox}%
      \addtocounter{totalfindings}{1}
    }
    
  }
{
    \newcommand{\finding}[1]{}
}

\ifthenelse{\boolean{showboxes}}
{
	\newcommand{\rqanswer}[1]{%
		\begin{tcolorbox}[enhanced,skin=enhancedmiddle,borderline={1mm}{0mm}{MidnightBlue},boxsep=3pt]
			\small
			#1
		\end{tcolorbox} 
    }
}
{
	\newcommand{\rqanswer}[1]{}
}

\makeatletter

\newcommand{\EulerC}{{\sc Euler}\xspace}

\newcommand{\tool}{{\sc AstroBR}\xspace}

\definecolor{mycolor}{RGB}{204,204,204}

\newcounter{myboxcounter}

\setboolean{showcomments}{true}

\ifthenelse{\boolean{showcomments}}
  {\newcommand{\nb}[2]{
    \fbox{\bfseries\sffamily\scriptsize#1}
    {\sf\small$\blacktriangleright$\textit{#2}$\blacktriangleleft$}
   }
   
  }
  {\newcommand{\nb}[2]{}
   
  }

\usepackage{changepage,threeparttable} 
\usepackage{hhline}

	\newcommand{\lucene}{{\scshape{Lucene}}\xspace}
	\newcommand{\sbert}{{\scshape{SBert}}\xspace}
	\newcommand{\clip}{{\scshape{Clip}}\xspace}
	\newcommand{\blip}{{\scshape{Blip}}\xspace}

	\newcolumntype{M}[1]{>{\centering\arraybackslash}m{#1}}

\newcommand{\svc}{{\scshape{Support Vector Machines}}\xspace}

\newcommand{\embgpt}{{\scshape{Gpt$_{emb}$}}\xspace}

\newcommand{\llamaft}{{\scshape{Llama-3$_{ft}$}}\xspace}

\begin{document}

\title{Studying and Automating Issue Resolution for Software Quality}

\author{Antu Saha}
\orcid{0009-0004-8656-7581}
\affiliation{%
  \institution{William \& Mary}
  \city{Williamsburg}
  \state{Virginia}
  \country{USA}
}
\email{asaha02@wm.edu}


\begin{abstract}

Effective issue resolution is crucial for maintaining software quality. Yet developers frequently encounter challenges such as low-quality issue reports, limited understanding of real-world workflows, and a lack of automated support. This research aims to address these challenges through three complementary directions. First, we enhance issue report quality by proposing techniques that leverage LLM reasoning and application-specific information. Second, we empirically characterize developer workflows in both traditional and AI-augmented systems. Third, we automate cognitively demanding resolution tasks, including buggy UI localization and solution identification, through ML, DL, and LLM-based approaches. Together, our work delivers empirical insights, practical tools, and automated methods to advance AI-driven issue resolution, supporting more maintainable and high-quality software systems.
\looseness=-1

\end{abstract}

\maketitle

\section{Problem and Research Statement}

During software maintenance, developers are tasked with resolving software issues, including defects, feature requests, enhancements, and other change requests~\cite{rajlich2011software,zeller2009programs}. This process—known as \textit{issue resolution}—requires diagnosing the reported problem, identifying its root cause, and implementing a fix. Effective issue resolution is essential for maintaining software quality: delays, misunderstandings, or incorrect fixes accumulate technical debt, increase maintenance effort, and reduce system reliability~\cite{ramasubbu2016technical}. As maintenance is often the most resource-intensive phase of the software lifecycle, efficient and accurate issue resolution is indispensable for sustaining system evolution and overall software quality~\cite{dehaghani2013factors}.
\looseness= -1

However, achieving effective issue resolution in practice remains challenging. User-submitted reports frequently contain incomplete or ambiguous information and often omit essential details, resulting in non-reproducible or misunderstood issues~\cite{Chaparro2017}. In addition, although prior work characterizes issue resolution as a linear sequence of stages~\cite{zhang2016literature,rajlich2011software}, how developers navigate and adapt these stages in real-world settings is not well understood. Finally, several core tasks in issue resolution remain highly manual and cognitively demanding; for example, identifying faulty UI elements from issue descriptions is time-consuming and susceptible to ambiguity~\cite{mahmud2024using}.
\looseness= -1

To address these challenges, our research enhances the issue resolution lifecycle across three complementary directions. First, we improve issue report quality by developing techniques that assess and enhance issue report elements (\eg reproduction steps) using LLM reasoning and application-specific information. Second, we investigate how developers resolve issues by empirically characterizing real-world workflows in both traditional and AI/ML-integrated systems. Third, we automate different resolution tasks—such as Buggy UI Localization and Solution Identification—through traditional ML, DL, and LLM-based approaches. 

Collectively, this work aims to produce new techniques, empirical insights, and practical tools that improve the effectiveness and efficiency of issue resolution. We propose automated methods for generating higher-quality issue reports, evidence-based models of real-world resolution practices, and data-driven automation for cognitively demanding tasks. These contributions support a more reliable, maintainable, and AI-driven issue resolution ecosystem, helping developers sustain software quality throughout system evolution.
\looseness= -1

\section{Proposed Research}
The ultimate goal of our research is to study and improve the issue resolution in practice to ensure software quality. We envision a future where both users and developers receive intelligent, actionable support throughout the entire issue resolution lifecycle—from reporting and understanding issues to diagnosing and resolving them. To realize this goal, our research advances three complementary directions: (1) improving issue report quality, (2) studying issue resolution workflows, and (3) automating issue resolution tasks.

\subsection{Improving Issue Report Quality}

High-quality issue reports are critical for effective issue resolution, as they enable developers to understand, reproduce, and ultimately resolve software problems. A well-written report should include three key components: the \emph{Observed Behavior (OB)}, the \emph{Expected Behavior (EB)}, and the \emph{Steps to Reproduce (S2Rs)}~\cite{Zimmermann2010, Bettenburg2008GoodBR}. However, issue reports often contain incomplete/ambiguous information or miss critical components~\cite{Chaparro2017}. Low-quality reports make it difficult for developers to reproduce and diagnose issues, increasing resolution time and the risk of incorrect or reopened fixes~\cite{Guo2010, Zimmermann2012}. Improving the clarity and completeness of OB, EB, and S2R descriptions is therefore essential for streamlining the issue resolution process.
\looseness=-1

\subsubsection{Assessing the Quality of Steps-to-Reproduce (S2R)}

To assess and improve S2R quality, we propose \tool~\cite{mahmud2025combining}, which combines large language models (LLMs) with dynamic application analysis to automatically evaluate and enhance S2R descriptions. \tool leverages GPT-4 to identify S2R sentences from an issue report, extract the atomic S2Rs, and map them with the application interactions from the app execution model constructed via dynamic analysis. By comparing the extracted S2Rs against the execution model, \tool detects correct, ambiguous, or missing steps and generates a quality report highlighting the quality issues. It also suggests missing steps that are required but not reported in the issue.
\looseness=-1


Our empirical evaluation on 21 bug reports from five Android applications shows that \tool outperforms the state-of-the-art technique, \ie~\EulerC~\cite{chaparro2019assessing}, improving S2R quality annotation performance by 25.2\% and missing step identification by 71.4\%  in terms of F1 scores. These results demonstrate that combining LLM reasoning with program execution data can enhance the assessment and improvement of S2Rs. The generated quality reports provide actionable feedback by pinpointing ambiguous or incomplete steps and suggesting missing steps, thereby enabling developers to reproduce issues more efficiently and accelerate their resolution.

\subsubsection{Improving the Quality of Issue Reports}

Prior research has proposed interactive reporting tools~\cite{song2022burt, Moran:FSE15} and LLM-based approaches~\cite{acharya2025can,Bo2024} to generate high-quality issue reports. However, these methods often rely solely on textual information presented in the issue report and lack application-specific information (\eg GUI interactions and application screen metadata), limiting their ability to produce accurate and reliable issue reports.

To address these limitations, we propose an approach that automatically \emph{enhances existing issue reports} by generating detailed, accurate, and complete OB, EB, and S2R descriptions. Our \textbf{key insight} is that LLMs can generate more precise reports when provided with app-specific information aligned with each issue report component. In particular, we envision utilizing four sources of information: (1) the original issue report, (2) the GUI interactions that capture app usage, (3) the app screens where those interactions occur, and (4)  the suspected buggy screen based on the reported OB. By constructing an execution model comprising screens and interactions and reasoning over it using LLMs, our proposed approach will generate high-quality issue reports.

We plan to evaluate this approach using a diverse set of Android issue reports and corresponding applications. The evaluation will extend existing S2R quality models~\cite{mahmud2025combining,chaparro2019assessing} into a unified framework assessing the correctness, completeness, and clarity of all three components, \ie OB, EB, and S2Rs. Generated reports will be evaluated by comparing them to the ground truth, the original user-submitted issue reports, and state-of-the-art LLM-based baselines~\cite{acharya2025can, mahmud2025combining}, using a combination of quantitative metrics and qualitative analysis. 
\looseness=-1


\subsection{Studying Issue Resolution Workflow}


Issue resolution is a critical sub-process within issue management, focused on diagnosing and fixing reported software problems. 
While the literature generally describes issue resolution as a linear sequence of steps—such as reproduction, analysis, solution design, implementation, and validation~\cite{zeller2009programs,rajlich2011software,eren2023analyzing}—there is limited understanding of how this process is implemented in practice, particularly how developers adapt or deviate from this idealized workflow. 
Understanding the practical workflow is essential for identifying bottlenecks, uncovering deviations from prescribed workflows, and providing developers with effective guidance. By examining real-world practices, we can also reveal recurring patterns that streamline future problem-solving and ultimately strengthen software maintenance. 
In this work, we aim to empirically characterize issue resolution workflows in both traditional and AI/ML-integrated systems. 
\looseness=-1

\subsubsection{Issue Resolution Workflow of Traditional Systems}

To understand how developers resolve software issues in practice, we conduct an empirical study~\cite{saha2025decoding} on the Mozilla Firefox. Our goal is to uncover the issue resolution stages, design an overall issue resolution process model, and derive the recurrent workflows (\ie patterns) adopted by developers for resolving issues. Using a mixed-methods approach, we qualitatively and quantitatively analyze the discussions contained in 356 issue reports from Mozilla Firefox. 
\looseness=-1

We identified six core stages of issue resolution—reproduction, analysis, solution design, implementation, code review, and verification—and used them to construct Mozilla’s overall issue resolution process model. Our analysis shows that, in practice, the process is \emph{highly iterative and non-linear}, with developers frequently revisiting between implementation, review, and verification. Moreover, we derived a catalog of 47 recurring resolution patterns that capture common instances of the process model. 
Our findings highlight the need for tools and methodologies that effectively support dynamic, feedback-driven workflows. The observed patterns and process insights can train new developers, inform process improvements, and provide a foundation for future automated support.



\subsubsection{Issue Resolution Workflow of AI/ML-Integrated Systems}

The growing integration of AI/ML components into software systems introduces new complexity to issue resolution due to the unique characteristics of such systems, \eg non-deterministic behavior, data dependencies, and evolving model versions. However, how these factors affect resolution, the challenges developers face, and their strategies to tackle them remain largely unexplored. To address this gap, we will conduct an empirical study to characterize issue resolution in AI/ML-integrated systems. Our goal is to understand (1) the unique activities involved, (2) the challenges developers face, and (3) the best practices they follow. We will start with a comprehensive literature review and analyze a representative sample of AI/ML-related issues using the annotation framework from our prior study~\cite{saha2025decoding}. Based on these findings, we will design a survey—and possibly follow-up interviews—to gather detailed insights from AI/ML developers about their real-world practices.

We hypothesize that issue resolution in AI/ML systems involves distinct activities (\eg dataset validation, model retraining, and performance evaluation), which introduce challenges around data quality, model interpretability, and tool integration. To test this, we will analyze survey responses using mixed methods, combining qualitative thematic analysis with quantitative summaries. We will then compare these results to our prior findings on traditional systems to highlight key similarities, differences, and areas for improvement.
\looseness=-1

\subsection{Automating Issue Resolution Tasks}


While recent advances have automated various aspects of issue resolution—such as bug reproduction~\cite{feng2022gifdroid,zhang2023automatically} and code localization~\cite{saha2024toward,mahmud2024using,Adnan:msr25}—several critical tasks still require substantial manual effort and remain underexplored. To understand a reported problem, developers must identify the faulty UI screens and components where the issue manifests, and to reuse past knowledge, they must manually locate solution-related discussions within issue reports. 
For mobile apps, where most issues are visually observable, developers must reason bug descriptions and connect them to visual UI representations—a challenging task known as \textbf{Buggy UI Localization}~\cite{saha2024toward}. Likewise, during resolution, developers generate rich \textit{solution-related content} that is often buried in long discussions, and automatic \textbf{Solution Identification}~\cite{saha2025automatically} can help them understand prior fixes, prevent regressions, and accelerate future resolutions.

\subsubsection{Buggy UI Localization}


To assess the feasibility of automating buggy UI localization, we conduct the first empirical study~\cite{saha2024toward} that evaluated whether existing deep learning (DL) and information retrieval (IR) techniques can automatically map a bug description to its corresponding UI elements. We formulate the problem as a retrieval task, where the observed behavior (OB) from an issue report serves as a query to identify relevant UI screens (\ie \textit{screen localization}) and UI components (\ie \textit{component localization}). We evaluate four techniques: one unsupervised text-based model (\lucene) and three supervised DL models—a textual model (\sbert) and two multi-modal models (\clip, \blip)—using textual and visual information extracted from real Android applications.

We found that while the multi-modal model, \blip, performs best for screen localization, the textual model, \sbert, excels for component localization. The best-performing approaches suggest the correct buggy UI screen and UI component within the top-3 recommendations for 52\% and 60\% of the cases, respectively. These results demonstrate the feasibility of leveraging existing language–vision models for Buggy UI Localization and highlight the complementary strengths of textual and visual representations. We further show that buggy UI localization benefits downstream tasks: integrating automatically identified buggy screens into traditional \textbf{buggy code localization} yields a 9–12\% improvement in Hits@10 over baselines that do not use UI information.


\looseness=-1

\subsubsection{Solution Identification}



To automate solution identification, we conduct an empirical study~\cite{saha2025automatically} leveraging 12 classifiers across three categories: six traditional machine learning models (MLMs), four pre-trained language models (PLMs), and two large language models (LLMs). We examined three application paradigms of language models—\textit{embeddings}, \textit{prompting}, and \textit{fine-tuning}—using a dataset of 356 Mozilla Firefox issues containing 4,867 comments. While prior studies have explored traditional ML classifiers for similar tasks~\cite{arya2019analysis,shi2021ispy}, it remains unclear how different model families— MLMs, PLMs, and LLMs—perform for solution identification.

Our results show that fine-tuned LLMs outperform all other models, with \llamaft achieving an F1 score of 0.716, while ensembles of top-performing models from each type further improve performance to 0.737 F1. Interestingly, lightweight classifiers using LLM embeddings (\eg\ \svc + \embgpt) perform competitively, offering a practical alternative for projects with limited computational resources. Qualitative analysis revealed that misclassifications often stem from missing conversational context or misleading textual clues, underscoring the need for context-aware approaches that analyze comment threads holistically. Additionally, our generalizability study shows that models trained on Mozilla can effectively transfer to other projects, achieving further performance gains with only a small amount of project-specific data.

\section{Anticipated Contributions}



Our research aims to improve the effectiveness and efficiency of software issue resolution, ensuring high software quality. To realize AI-driven issue resolution, we support both users and developers by improving issue report quality, understanding resolution workflows in practice, and automating key resolution tasks. First, we improve issue report quality with automated, context-aware techniques that assess and refine Observed Behavior, Expected Behavior, and Steps to Reproduce. Next, we empirically analyze developer workflows in traditional and AI/ML-integrated systems, uncovering real practices and challenges. Finally, we automate critical tasks, including Buggy UI Localization and Solution Identification, using data-driven and LLM-based approaches. Together, these contributions provide new techniques, empirical insights, and open-source tools to enable a more intelligent, reliable, and maintainable issue resolution ecosystem.


\bibliographystyle{ACM-Reference-Format}
\bibliography{references}

\end{document}